\documentclass[spanish]{book}
\usepackage{graphicx}
\usepackage[intlimits]{amsmath}
\usepackage[spanish]{babel}
\usepackage{amsfonts}
\usepackage{amssymb}
\newtheorem{theorem}{Teorema}

\newtheorem{axiom}{Axioma}

\newtheorem{definition}{Definición}

\newtheorem{notation}{Nota}

\begin{document}

\frontmatter
\title{Aproximaci\'{o}n Discreta de la Relatividad General}
\author{Eduard Alexis Larra\~{n}aga Rubio
\and Universidad Nacional de Colombia}
\maketitle
\tableofcontents

\pagebreak \ 

\section{\bigskip INTRODUCCION}

Einstein nos present\'{o} hace ya casi 100 a\~{n}os una descripci\'{o}n de la
gravedad como geometr\'{i}a. Ahora, el C\'{a}lculo de Regge nos presenta una
aproximaci\'{o}n a la Relatividad General en la que la variedad Espacio-Tiempo
es modelada mediante bloques 4-dimensionales con geometr\'{i}a plana (
Lorentziana ) llamados \textit{Simplex}. M\'{a}s exactamente, un complex
simplicial geom\'{e}trico cuadridimensional es utilizado como un modelo del espacio-tiempo.

\bigskip

Este conjunto de notas es una peque\~{n}a introducci\'{o}n al m\'{e}todo
desarrollado por Regge$^{\cite{regge}}$ y su utilizaci\'{o}n en la
soluci\'{o}n de problemas en Relatividad General.

En este modelo, las propiedades geom\'{e}tricas de la variedad son expresadas
en t\'{e}rminos de la longitud de las aristas del complex. De esta manera, el
C\'{a}lculo de Regge se convierte en una teor\'{i}a discretizada en la que las
longitudes de las aristas son las variables din\'{a}micas fundamentales.

\bigskip

Supongamos ahora que conocemos la geometr\'{i}a de una hipersuperficie
como-de-espacio de la variedad 4-dimensional espacio-tiempo. Ya que las
ecuaciones geometrodin\'{a}micas de Einstein son deterministas, la
geometr\'{i}a a partir de este instante estar\'{a} completamente predeterminada.

En t\'{e}rminos del C\'{a}lculo de Regge, esto quiere decir que sera posible
calcular la longitud de las aristas de los nuevos simplex a partir de las
longitudes de los simplex anteriores. Una vez se halla discretizado o
esqueletizado toda la variedad, tendremos un ''cat\'{a}logo'' con la longitud
de todas las aristas o huesos, y con ello conoceremos la geometr\'{i}a
completa de cada hipersuperficie.

\bigskip

\mainmatter

\part{C\'{a}lculo de Regge}

\chapter{\bigskip Simplex, Complex Simpliciales y Poliedros}

\bigskip Comenazaremos definiendo los bloques fundamentales con los que se
modelar\'{a} una variedad n-dimensional.

\begin{definition}
\bigskip Dados $n+1$ puntos en $\mathbb{R}^{m}$, $(m\geq n)$, enumerados por
$p_{0,}p_{1},p_{2},...,p_{n}$, si estos puntos son geom\'{e}tricamente
independientes (i.e no existe un hiperplano (n-1)-dimensional que contenga a
los n+1 puntos) , entonces definimos el simplex n-dimensional $T_{n}%
=<p_{0}p_{1}....p_{n}>$ como :%

\begin{equation}
T_{n}=\left\{  x\in\mathbb{R}^{m}:x=\underset{i=1}{\overset{n+1}{\sum}}%
c_{i}p_{i}\text{ \ \ , \ donde \ }c_{i}\geq0\text{ \ \ y \ \ }\underset
{i=1}{\overset{n+1}{\sum}}c_{i}=0\text{ \ }\right\}
\end{equation}
\end{definition}

\bigskip

\ As\'{i}, tenemos que un 0-simplex $T_{0}=<p_{0}>$ es un punto o v\'{e}rtice,
un 1-simplex $T_{1}=<p_{1}p_{2}>$ es una linea o arista, un 2-simplex
$\ T_{2}=<p_{0}p_{1}p_{2}>$ es un triangulo con su interior incluido y un
3-simplex $T_{3}=<p_{0}p_{1}p_{2}p_{3}>$ es un tetrahedro s\'{o}lido.

\begin{notation}
Ya que $T_{n}$ es un subconjunto cerrado de $\mathbb{R}^{m}$ entonces $T_{n}$
es compacto.
\end{notation}

\begin{definition}
Sea $T_{n}=<p_{0}p_{1}...p_{n}>$ un n-simplex y $0\leq r\leq n$ un entero. Si
tomamos $r+1$ puntos $p_{i_{0}},p_{i_{1}},...,p_{i_{r}}$ \ del conjunto
$p_{0},p_{1},...,p_{n}$ entonces estos forman un r-simplex $T_{r}=<p_{i_{0}%
}p_{i_{1}}...p_{i_{r}}>$ y se denomina una r-cara del simplex $T_{n}$.
\end{definition}

\begin{notation}
El numero de r-caras de un n-Simplex es $\binom{n+1}{r+1}$
\end{notation}

\begin{definition}
Sea $K$ un conjunto finito de simplex en $\mathbb{R}^{n}$. Si estos simplex
estan unidos entre si y satisfacen:

\begin{enumerate}
\item
\begin{enumerate}
\item
\begin{enumerate}
\item \ Si $\sigma$ es una cara de un simplex en $K,$ entonces $\sigma\in K$

\item  Si $T$ y $T%
\acute{}%
$ son simplex de $K$, entonces $T\cap T%
\acute{}%
$ es vacio o una cara de $T$ y de $T%
\acute{}%
$
\end{enumerate}
\end{enumerate}
\end{enumerate}

entonces $K$ se denomina un Complex Simplicial.
\end{definition}

\begin{notation}
La dimensi\'{o}n de un Complex $K$ se define como la m\'{a}xima dimensi\'{o}n
de los simplex del conjunto $K.$
\end{notation}

\begin{definition}
Sea $K$ un Complex. Si cada simplex perteneciente a $K$ es considerado como un
subconjunto de $\mathbb{R}^{n}$ $(n\geq\dim K)$ entonces la uni\'{o}n de todos
los simplex es un subconjunto de $\mathbb{R}^{n}$. Este subconjunto se
denomina el Poliedro $\left|  K\right|  $ del complex simplicial $K$.
\end{definition}

\begin{notation}
La dimensi\'{o}n de $\left|  K\right|  $ como subconjunto de $\mathbb{R}^{n}$
es la misma de $K$. $(\dim\left|  K\right|  =\dim K).$
\end{notation}

\bigskip

\begin{definition}
Sea $X$ un espacio Topol\'{o}gico. Si existe un Complex $K$ y un homeomorfismo
$f:\left|  K\right|  \longrightarrow X$, entonces se dice que $X$ es
triangulable y el par $(K,f)$ se denomina una Triangulaci\'{o}n de $X$.
\end{definition}

\bigskip

\section{Poliedros y variedades bi-dimensionales}

Comenzaremos ahora con una descripci\'{o}n inicial de la teor\'{i}a de
curvatura intr\'{i}nseca en poliedros en un caso simple tal como lo es una
variedad bidimensional (superficie) para luego realizar su generalizaci\'{o}n
a $n$ dimensiones. \ En general una superficie arbitraria puede ser
considerada como el l\'{i}mite de una sucesi\'{o}n de poliedros cuyo
n\'{u}mero de caras va incrementandose. La aproximaci\'{o}n que nos ofrecen
estos poliedros es mala si se observa de cerca la superficie, pero para un
observador lejano esta es satisfactoria.

\subsection{Curvatura Integral Gaussiana}

En cualquier superficie puede definirse una curvatura integral Gaussiana
mediante el uso de triangulos construidos con geod\'{e}sicas.

Supongamos un triangulo $t$ cuyos angulos internos son $\ \alpha,\beta,\gamma
$. En general, para geometr\'{i}as no-Euclideanas tenemos $\alpha+\beta
+\gamma\neq\pi$. As\'{i}, definimos la \textit{Curvatura Integral Gaussiana
}$\varepsilon_{Gt}$\textit{\ }del triangulo $t$ como:%

\begin{equation}
\varepsilon_{Gt}=a+\beta+\gamma-\pi
\end{equation}

De esta manera, por ejemplo, en una esfera, la curvatura integral Gaussiana
ser\'{a}: $\varepsilon_{Gt}=\frac{A_{t}}{R^{2}}$ donde $A_{t}$ es el area del
triangulo $t$ y $R$ es el radio de la esfera.

\bigskip

Ahora bien, si $t$ se reduce hasta convertirse un punto $\ P$ y el l\'{i}mite
$\underset{A\rightarrow0}{\lim}\frac{\varepsilon_{Gt}}{A_{t}}=K(P)$ existe y
es independiente de la forma de realizarse, se toma la funci\'{o}n $K(P)$
\-como la definicion de \textit{Curvatura Gaussiana Local en el punto }%
$P$\textit{. }

\bigskip

En el caso de la esfera tenemos $K(P)=\frac{1}{R^{2}}$ .

De esta manera tenemos que la Curvatura Integral Gaussiana se puede escribir como:%

\begin{equation}
\ \ \varepsilon_{Gt}=\int\limits_{\tau}K(P)dA
\end{equation}

Sin embargo debemos hacer claridad en que la curvatura Integral puede existir
bajo pocas suposiciones, mientras que la curvatura local puede no existir como
una funci\'{o}n ordinaria, sino como una medida.

\bigskip

Si consideramos ahora un poliedro $M$, observamos que si tomamos un triangulo
$t$ ubicado completamente en una de las caras del poliedro, tendremos
$\varepsilon_{Gt}=0$. Este mismo resultado se obtiene si el triangulo no
incluye ninguno de los vertices de $M$. De aqui concluimos que $K(P)=0$ si $P$
no es un v\'{e}rtice.

Por otro lado, si $t$ incluye el vertice $V$ de $M$, la curvatura integral no
depender\'{a} de la forma de $t$. Es decir que $\varepsilon_{Gt}%
=\varepsilon_{V}$ solo depende del v\'{e}rtice que se toma. Para encontrar la
\textit{deficiencia} del vertice $V$ tomamos la suma de todos los \'{a}ngulos
internos de las caras de $M$ con v\'{e}rtice $V$, y esta debe valer
$2\pi-\varepsilon_{v}$.

En general si $t$ contiene varios v\'{e}rtices $V,W,X,$etc la curvatura
ser\'{a} $\varepsilon_{Gt}=\varepsilon_{V}+\varepsilon_{W}+\varepsilon
_{X}+...$ y por lo tanto $K(P)$ resulta ser una funci\'{o}n Delta de Dirac con
los v\'{e}rtices como soporte.

\bigskip

\subsubsection{Formula De Euler para Poliedros}

Para observar un ejemplo del tratamiento de la curvatura, consideremos un
poliedro $M$ compacto (i.e. finito y cerrado). El teorema para la curvatura
integral de Gauss-Bonnet se puede escribir como:%

\begin{equation}
\underset{n}{\sum}\varepsilon_{n}=2\pi(2-N)
\end{equation}

donde $N$ es la clase de $M$ ( e.g. $N=0$ para la esfera, $N=2$ para el toro).
Ahora bien, si suponemos, sin p\'{e}rdida de generalidad, que las caras de $M$
son tri\'{a}ngulos, y denotamos por $\sigma_{fn}$ el \'{a}ngulo interno de la
cara $f$ con v\'{e}rtice $n$, tenemos por la definici\'{o}n de la secci\'{o}n anterior:%

\begin{equation}
\underset{f}{\sum}\sigma_{fn}=2\pi-\varepsilon_{n}%
\end{equation}

donde la sumatoria se realiza sobre las caras $f$ que tengan a $n$ como
v\'{e}rtice. Tenemos adem\'{a}s que%

\begin{equation}
\underset{n}{\sum}\sigma_{fn}=\pi
\end{equation}

donde la sumatoria se realiza ahora sobre los v\'{e}rtices de la cara $f$.

\bigskip

Ahora bien, la sumatoria doble $\underset{fn}{\sum}\sigma_{fn}$ puede ser
obtenida de dos maneras diferentes: Sumando primero sobre los v\'{e}rtices y
luego sobre las caras tenemos:%

\begin{equation}
\bigskip\underset{fn}{\sum}\sigma_{fn}=\underset{f}{\sum}\underset{n}{\sum
}\sigma_{fn}=\underset{f}{\sum}\pi=\pi f
\end{equation}

donde $f$ es el n\'{u}mero de caras de $M$.

Por otro lado, sumando primero sobre las caras y utilizando el Teorema de
Guass-Bonnet tenemos:%

\begin{equation}
\underset{fn}{\sum}\sigma_{fn}=\underset{n}{\sum}\underset{f}{\sum}\sigma
_{fn}=\underset{n}{\sum}2\pi-\varepsilon_{n}=2\pi f-\underset{n}{\sum
}\varepsilon_{n}=2\pi V-2\pi(2-N)
\end{equation}

donde $V$ es el n\'{u}mero de V\'{e}rtices de $M$. Ya que estas dos dos
maneras de realizar la sumatoria deben llevar a resultados iguales tenemos:%

\begin{equation}
f=2(V-2+N)
\end{equation}

Ya que las caras de $M$ son tri\'{a}ngulos tenemos que $2e=3f$ donde $e$ es el
n\'{u}mero de aristas \ de $M$. Reemplazando este valor en la ecuaci\'{o}n
anterior obtenemos la formula de Euler para poliedros:%

\begin{equation}
V-e+f=2-N
\end{equation}

\section{\bigskip Matriz de Conexion y Transporte Paralelo}

Veamos ahora como una variedad $M$ puede ser modelada mediante una red
simplicial o complex. Una vez construida esta red, la geometr\'{i}a
intr\'{i}nseca de $M$ vendra dada por la longitud de sus aristas y por la
llamada \textit{Matriz de Conexion}, \ la cual es en escencia una lista de
todas las caras, aristas y v\'{e}rtices de $M$ junto con su relaci\'{o}n
mutua. Esta matriz nos permitir\'{a} conocer cuales v\'{e}rtices y aristas
pertenecen a una cara dada, y por lo tanto en ella esta impl\'{i}cita toda la
informacion topol\'{o}gica de $M$. Adem\'{a}s, al conocer la longitud de todas
las aristas, tendremos tambi\'{e}n las deficiencias $\varepsilon_{n}$ de cada
v\'{e}rtice.

De esta manera, la noci\'{o}n de red simplicial reemplaza la red de
coordenadas que se utilizan en la variedad, mientras que el tensor m\'{e}trico
ser\'{a} reemplazado por la longitud de las aristas.

\bigskip

Cuando el n\'{u}mero de v\'{e}rtices y/o caras de $M$ se incrementa, la
Curvatura Local se puede aproximar a una funci\'{o}n continua: $K(P)=\varrho
\varepsilon$ , donde $\varrho$ es la densidad de v\'{e}rtices y $\varepsilon$
su deficiencia. Esta aproximaci\'{o}n es valida en el caso en que este
producto varie suavemente dentro del tri\'{a}ngulo $t.$

Al trabajar en dimensiones mas altas es conveniente utilizar el Transporte
Paralelo de Levi-Civita \ (TP) para medir la curvatura.

Supongamos dos puntos $P$ y $Q$ de $M$, y un arco $a$ que los une. Entonces,
el TP es un mapeo ortogonal del espacio vectorial $S_{p}$ en el punto $P$ al
espacio vectorial $S_{q}$ en el punto $Q$. En particular, si $P=Q$ tenemos que
$a$ es un aro y el TP mapea $S_{p}$ en si mismo, por lo tanto TP debe ser una
rotaci\'{o}n de $S_{p}$ en un \'{a}ngulo $\varepsilon(a)$ que depende del
camino recorrido $a$. Este \'{a}ngulo corresponde precisamente a la curvatura
integral Gaussiana:%

\begin{equation}
\varepsilon(a)=\underset{a}{\int}K(P)dA
\end{equation}

ya que si por ejemplo la curva cerrada $a$ corresponde a un trangulo $t$
tendremos $\varepsilon(a)=\varepsilon_{Gt}.$

Adem\'{a}s, $\varepsilon$ es una funci\'{o}n aditiva en el camino que se
recorre. Si tomamos dos aros $a$ y $b$ con el mismo punto final $P$ y
realizamos un TP alrededor del ''producto'' $ab$ (i.e. recorremos primero el
aro $a$ y luego continuamos por el aro $b$), \ tenemos:%

\begin{equation}
\varepsilon(ab)=\varepsilon(a)+\varepsilon(b)
\end{equation}

No consideraremos casos en los cuales un v\'{e}rtice de $M$ pertenece al aro
$a$, ya que en este caso el TP no se puede definir sin ambig\"{u}edades.

Si un aro $a$ puede ser deformado continuamente hasta obtener un aro $b$,
manteniendo $P$ fijo y sin encontrarnos con un v\'{e}rtice, tenemos que
$\varepsilon(a)=\varepsilon(b)$ y por lo tanto escribiremos $a=b$.

\bigskip

\subsection{$\varepsilon-$Conos}

Consideremos ahora una variedad $M$ con un solo v\'{e}rtice.Esta variedad
puede imaginarse entonces como un cono o una piramide. Tomando coordenadas
polares planas $\rho$ y $\theta$ \ con origen en el v\'{e}rtice, la
m\'{e}trica vendr\'{a} dada por:%

\begin{equation}
ds^{2}=d\rho^{2}+\rho^{2}d\theta^{2}%
\end{equation}

En el plano Euclideano dos puntos con coordenadas $\rho$ iguales y con
coordenadas $\theta$ diferentes en un multiplo de $2\pi$ son identicos. En la
variedad $M$ dos puntos con la misma coordenada $\rho$ ser\'{a}n identicos si
sus coordenadas $\theta$ difieren en un multiplo de $2\pi-\varepsilon.$

Esta variedad es conocida como el $\varepsilon-Cono$, y es Euclideana en todos
sus puntos excepto en el \'{e}rtice, el cual tiene una deficiencia
$\varepsilon.$

\bigskip

Consideremos ahora la recta real $R$ con variable $z$. El producto
$R\otimes\varepsilon-Cono$ es una variedad que posee la m\'{e}trica:%

\begin{equation}
ds^{2}=dz^{2}+d\rho^{2}+\rho^{2}d\theta^{2}%
\end{equation}

Este espacio es Euclideano excepto en la linea $\rho=0$ y es conocido como
$\varepsilon-3Cono$

\bigskip

En general, si consideramos el producto $R^{n-2}\otimes\varepsilon-Cono$ con
la m\'{e}trica%

\begin{equation}
ds^{2}=dz_{1}^{2}+dz_{2}^{2}+...+dz_{n-2}^{2}+d\rho^{2}+\rho^{2}d\theta^{2}%
\end{equation}

obtenemos entonces un $\varepsilon-nCono$ y esta variedad es Euclideana
excepto en el subconjunto plano $n-2$ dimensional $\rho=0.$

\chapter{Variedades n-Dimensionales}

\section{El Grupo Fundamental}

Consideraremos ahora una Variedad Rimanniana \textit{n}-dimensional $M$.
Tenemos entonces las siguientes definiciones:

\bigskip

\begin{definition}
Una trayectoria $a$ es una funcion continua y univaluada $A(s)$ de un
par\'{a}metro $s\in\mathbb{R}$ , $0\leq s\leq1.$ Los puntos $A(0)$ y $A(1)$
son los puntos inicial y final de la trayectoria.

Si $A(0)=A(1)=A$ la trayectoria se denomina Aro
\end{definition}

\begin{definition}
Sean $a$ y $b$ dos trayectorias definidas por las funciones $A(s)$ y $B(s)$
tales que $A(1)=B(0)$. Entonces el producto $c=ab$ es definido por la
funci\'{o}n:

$C(s)=\left\{
\begin{array}
[c]{c}%
A(2s)\text{ \ \ \ \ \ \ \ \ \ \ para }s<0.5\\
B(2s-1)\text{ \ \ \ para }s\geq0.5
\end{array}
\right.  $

Este producto de trayectorias es asociativo pero en general no es conmutativo
($ab\neq ba$).

Si $a$ y $b$ son aros con el mismo punto final $\ P$ entonces $ab$ y $ba$
existen y son aros con punto final $P.$
\end{definition}

\begin{definition}
Sea $a$ una trayectoria definida por la funcion $A(s)$. La trayectoria inversa
$a^{-1}$ esta definida por la funcion $A(1-s)$.
\end{definition}

\bigskip

Consideraremos ahora que la variedad $M$ es Euclideana excepto en un
subconjunto cerrado $w\subset M$ que posee curvatura. Sean entonces $a$ y $b$
dos trayectorias con los mismos puntos iniciales y finales, y definidas por
las funciones $A(s)$ y $B(s)$. Ademas supondremos siempre que $A(s)\notin w$ y
$B(s)\notin w$. Definimos entonces:

\begin{definition}
Si existe una funcion univaluada $P(s,t)$ de $s\in\mathbb{R}$ , $0\leq s\leq1$
y de $t\in\mathbb{R}$ , $0\leq t\leq1$ continua en los par\'{a}metros $t$ y
$s$ , tal que $P(s,0)=A(s)$ y $P(s,1)=B(s)$ decimos que se ha deformado $a$ en
$b$ y la funci\'{o}n $P(s,t)$ se llama una deformaci\'{o}n.
\end{definition}

\begin{definition}
Sea una deformaci\'{o}n $P(s,t)$ que deforma $a$ en $b$ tal que $P(s,t)\notin
w$ para todo $s,t$. Escribimos entonces $a\sim b$.
\end{definition}

\begin{theorem}
El simbolo $\sim$ satisface las propiedades formales de una equivalencia
(holonom\'{i}a).
\end{theorem}

Para probar el anterior teorema sean $a$ y $b$ dos trayectorias.

Teneemos que $a\sim a$ ya que podemos tomar como deformaci\'{o}n la
funci\'{o}n $P(s,t)=A(s)$

Si $a\sim b$ entonces $P(s,1-t)$ deforma $b$ en $a$ y $P(s,1-t)\notin w$ por
lo tanto $b\sim a$

Por \'{u}ltimo, si $P(s,t)$ deforma $a$ en $b$ y $Q(s,t)$ deforma $b$ en $c$ y
ademas $a\sim b$ $,$ $b\sim c$, entonces la funci\'{o}n :

\bigskip

$R(s,t)=\left\{
\begin{array}
[c]{c}%
P(s,2t)\text{ \ \ \ \ \ \ \ \ \ para }t<0.5\\
Q(s,2t-1)\text{ \ \ para }t\geq0.5
\end{array}
\right.  $

\bigskip

deforma $a$ en $c$ y ademas $\ R(s,t)\notin w$ por lo que $a\sim c$.

\bigskip

Las anteriores definiciones se aplican de la misma forma a los aros con la
condici\'{o}n adicional de que despues de \ la transformaci\'{o}n la nueva
trayectoria sea un aro, esto es $P(0,t)=P(1,t)$ para todo $t$.

\bigskip

\begin{definition}
Sean $a$ y $b$ dos aros y $P(s,t)$ una deformaci\'{o}n que lleva $a$ en $b$.
Si se cumple que $P(0,t)=P(1,t)=P(0,0)=P$ \ \ entonces escribimos $a\approx b$.

El signo $\approx$ se aplica unicamente a aros. Claramente $a\approx b$
implica $a\sim b$ pero la implicaci\'{o}n inversa no es cierta.
\end{definition}

\begin{theorem}
El simbolo $\approx$ define una equivalencia (homotop\'{i}a \ en $M-w$)
\end{theorem}

Para probar este teorema sean $a$ y $b$ dos aros y $P(s,t)$ una deformaci\'{o}%
n que llea $a$ en $b$.

Observemos que $a\approx a$ ya que podemos tomar como deformaci\'{o}n la
funci\'{o}n $P(s,t)=A(s)$

Si $a\approx b$ entonces $P(s,1-t)$ deforma $b$ en $a$ y $P(s,1-t)\notin w$
por lo tanto $b\approx a$

Por \'{u}ltimo, si $a,b$ y $c$ son aros y $P(s,t)$ deforma $a$ en $b$ y
$Q(s,t)$ deforma $b$ en $c$ y ademas $a\approx b$ $,$ $b\approx c$, entonces
la funci\'{o}n :

\bigskip

$R(s,t)=\left\{
\begin{array}
[c]{c}%
P(s,2t)\text{ \ \ \ \ \ \ \ \ \ para }t<0.5\\
Q(s,2t-1)\text{ \ \ para }t\geq0.5
\end{array}
\right.  $

\bigskip

deforma $a$ en $c$ y ademas $\ R(s,t)\notin w$ por lo que $a\approx c$.

\bigskip

\begin{definition}
Defnimos ahora el aro unidad $u$ en un punto $P$ por medio de la funci\'{o}n $U(s)=P=const.$

Claramente para cualquier trayectoria $a$ tenemos $aa^{-1}=u$
\end{definition}

\begin{theorem}
Sean $a,a%
\acute{}%
,b$ y $b%
\acute{}%
$ aros tales que \ $a\approx a%
\acute{}%
$ y $b\approx b%
\acute{}%
$ . Entonces $ab\approx a%
\acute{}%
b%
\acute{}%
$.
\end{theorem}

\bigskip

Para la prueba del teorema sean $P(s,t)$ la funci\'{o}n que deforma $a$ en $a%
\acute{}%
$ y $Q(s,t)$ la funci\'{o}n que deforma $b$ en $b%
\acute{}%
$. Entonces la funci\'{o}n

\bigskip

$R(s,t)=\left\{
\begin{array}
[c]{c}%
P(2s,t)\text{ \ \ \ \ \ para }s<0.5\\
Q(2s-1)\text{ \ para }s\geq0.5
\end{array}
\right.  $

\bigskip

es una deformaci\'{o}n de $ab$ a $a%
\acute{}%
b%
\acute{}%
$ , con $R(0,t)=R(1,t)$ y adem\'{a}s $R(s,t)\notin w$ .

Por lo tanto tenemos $ab\approx a%
\acute{}%
b%
\acute{}%
$.

\bigskip

Si tomamos ahora un conjunto de aros con punto final $P$, podemos dividirlo en
clases de equivalencia. Cada una de estas clases esta caracterizada por uno de
sus elementos, y por lo tanto utilizaremos uno de sus aros para identificar la
clase completa.

Definimos entonces el \textit{GRUPO\ FUNDAMENTAL} como el conjunto de todas
las clases de equivalencia con el producto de trayectorias dado anteriormente.
Para demostrar que se cumplen los axiomas de un Grupo veamos que:

Si $a$ y $b$ son clases entonces todos los productos $a%
\acute{}%
b%
\acute{}%
$ con $a\approx a%
\acute{}%
$ y $b\approx b%
\acute{}%
$ son homot\'{o}picos y pertenecen a una nueva clase que definimos como el
producto $ab$. Este producto de clases es asociativo.

Si $a$ es una clase y $u$ es la clase unitaria entonces $au=ua=a$.

A cada clase $a$ asociamos una clase inversa $a^{-1}$ tal que $aa^{-1}%
=a^{-1}a=u$.

\bigskip

A cada aro $a$ se le puede asociar una matriz ortogonal $S(a)$ de la siguiente manera:

Tomamos un vector $\mathbf{A}$ en el punto $P$ y lo transportamos
paralelamente (TP) a traves de $a$ .Existe entonces un mapeo lineal y que
preserva la norma que relaciona las posiciones inicial y final del vector
$\mathbf{A}$. Si escogemos un sistema de referencia ortogonal local en $P$,
este mapeo se puede representar mediante una matriz ortogonal $S(a)$.

Como caso particular, si nuestro sistema de referencia es localmente
pseudo-Euclideano (e.g. Lorentziano), las matrices que represetan el TP son
matrices de Lorentz.

\bigskip

Ahora bien, dos aros homot\'{o}picos produciran la misma posici\'{o}n final
del vector transpoprtado paralelamente, y por lo tanto seran representados por
la misma matriz. Esto nos indica que la matriz ortogonal $S(a)$ es funci\'{o}n
de las clases de Equivalencia, lo que implica que el conjunto de matrices
ortogonales son una \textit{Representacion del Grupo Fundamental en el Espacio
Vectorial del punto }$P$\textit{. }

\textit{\ }Para comprobar esto, veamos primero que a todos los aros $v\approx
u$ les corresonde la matriz unidad. Para ello sea $T(s,t)$ la funci\'{o}n de
deformaci\'{o}n que lleva $v$ en $u$. El conjunto de todos los puntos $T(s,t)$
es una superficie simplemente conexa $\Sigma$. Podemos tomar entonces en
$\Sigma$ y en una vecindad de $\Sigma$ apropiada, un sistema de referencia
globalmente Euclideano, y asi el resultado de un TP a lo largo de $v$
calculado en este sistema de referencia es la identidad.

Ahora bien, sean $S(a),S(b)$,$S(c)$ son las matrices correspondientes a los
aros $a,b$,\thinspace$c$ \ y $\ c=ab$. Tomamos un vector $\mathbf{V}$ en $P$ y
lo transportamos a lo largo de $a$. Obtenemos entonces $S(a)\mathbf{V}$. Si
transportamos ahora este vector a lo largo de $b$ obtenenmos
$S(b)S(a)\mathbf{V}$. Este resultado debe ser igual a transportar el vector
original $\mathbf{V}$ a lo largo de $c$, es decir $S(c)\mathbf{V}%
=S(b)S(a)\mathbf{V}$. Ya que el vector $\mathbf{V}$\textbf{\ }es arbitrario
obtenemos $S(c)=S(b)S(a)$.

Por ultimo, si $a\approx a%
\acute{}%
$ entonces podemos escribir $a=va%
\acute{}%
$ donde $v\approx u$. Tenemos entonces que $S(a)=S(v)S(a%
\acute{}%
)$, pero ya que $S(v)=1$ obtenemos $S(a)=S(a%
\acute{}%
)$. Es decir que la matriz es funci\'{o}n de la clase de equivalencia.

\bigskip

\subsection{Grupo Fundamental para los $\varepsilon-Conos$}

Como un caso particular tomamos un aro en un $\varepsilon-Cono$. Podemos
entonces dar el \'{a}ngulo $\theta$ para cada punto en el aro como una
funcio\'{o}n de $s$. Podemos entonces definir sin p\'{e}rdida de generalidad
$\theta(0)=0$. De esta manera tenemos entonces $\theta(1)=N(2\pi-\varepsilon)$
donde $N$ nos da el n\'{u}mero de vueltas que el aro da a la linea $\rho=0$
(\textit{Hueso}) el cual coincide con el subconjunto $w$ definido
anteriormente. Dos aros con igual $N$ son homot\'{o}picos. De esta manera
llamaremos $a(N)$ la clase de equivalencia correspondiente.

El producto de dos clases de equivalencia ser\'{a} $\ a(N)a(M)=a(N+M)$. Es
decir que el Grupo fundamental es Isomorfo al grupo de los enteros bajo
adici\'{o}n.

\bigskip

Para encontrar la representacion del Grupo Fundamental en t\'{e}rminos de las
matrices Ortogonales $S(a(n))=S(n)$ notemos que si llamamos $S(0)=S$ tenemos
que para un $N$ dado $S(N)=S^{N}$.

$S$ es llamdo el Generador del Hueso.

\bigskip

Consideremos ahora un $\varepsilon-nCono$. Tomamos un vector $V$ en el punto
$P$, y lo transportamos a lo largo de $a(1)$. E lvector $V$ puede expresarse
en t\'{e}rminos de dos componentes ortogonales, una en el $\varepsilon-Cono$ y
otra en $R^{n-2}$. La primera componente ser\'{a} rotada en un \'{a}ngulo
$\varepsilon$, mientras que la segunda permanece invariante. Esto nos muestra
que la representaci\'{o}n del Grupo Fundamental en este caso es una matriz
$n\times n$ con un espacio invariante $R^{n-2}$. por ejemplo en el caso de un
$\varepsilon-3Cono$ el espacio invariante es una Recta y por lo tanto la
matriz $S$ es una rotaci\'{o}n con eje $R^{n-2}=R$.

\bigskip

\section{\bigskip Espacios Esqueleto}

Los \textit{Espacios Esqueleto} corresponden a la generalizaci\'{o}n
$n$-dimensional de los Poliedros definidos anteriormente. En ellos la
curvatura se concentra en \ subespacios de dimension $n-2$ \ llamados
\textit{Huesos}. Una de estas estructuras son los $\varepsilon-nConos$.

Al igual que en el caso bidimensional comenzamos por una descomposici\'{o}n
simplicial de la variedad, en este caso $n$-dimensional $M$. Esta
descomposici\'{o}n determina la topolog\'{i}a de la variedad, pero no su
m\'{e}trica, la cual ser\'{a} definida siguiendo los siguientes axiomas:

\begin{axiom}
\bigskip La m\'{e}trica en el interior de cualquier simplex cerrado
n-dimensional $T_{n}$ es Plana (e.g. Euclideana en $M_{3}$ o Lorentziana en
$M_{4}$).

Podemos entonces calcular la distancia entre dos puntos cualquiera dentro del
Simplex, definir un sistema de coordenadas cartesiano y expresar las
coordenadas de los puntos de la frontera de $T_{n}$ en este sistema.
\end{axiom}

\begin{axiom}
En la m\'{e}trica propia de $T_{n}$, la frontera del Simplex se puede expresar
como la suma de $n+1$ simplex cerrados $T_{n-1}$ y estos simplex son planos.
\end{axiom}

\begin{axiom}
Si el simplex $T$ $_{n-1}$ es frontera com\'{u}n de $T_{n}$ y de $T%
\acute{}%
_{n}$ entonces la distancia entre dos puntos cualquiera de $T_{n-1}$ es la
misma en los sitemas de referencia de $T_{n}$ y de $T%
\acute{}%
_{n}$.
\end{axiom}

\begin{axiom}
Si $P\in T_{n}$ y $P%
\acute{}%
\in T%
\acute{}%
_{n}$ \ y $P,P%
\acute{}%
$ estan lo suficientemente cerca de la frontera comun $T_{n-1}$ entonces la
distancia $PP%
\acute{}%
$ se define como el infimo de las distancias $PQ+QP%
\acute{}%
$ para $Q\in T_{n-1}$.
\end{axiom}

\bigskip

Con estos axiomas obtenemos una m\'{e}trica plana tambi\'{e}n en $T_{n-1}$
excepto en las fronteras $T_{n-2}$. Adem\'{a}s esta definici\'{o}n es
estrictamente local, pero nos permite unir suavemente los simplex vecinos.

De esta manera, la Variedad $M$ es Plana excepto en la uni\'{o}n $w$ de los
simplex frontera o Huesos $T_{n-1}$.

\bigskip

En particular en una variedad tridimensional $M_{3}$ los Huesos son segmentos
de recta mientras que los $T_{n-3}$ son puntos.Cada hueso conecta dos
$T_{n-3}$. Si conocemos la longitud de todos y cada uno de los huesos,
entonces podemos conocer la estructura geom\'{e}trica de los simplex y de la
variedad. En particular podemos conocer tambi\'{e}n el \'{a}ngulo dihedral.
Por ejemplo, si en un hueso dado no existe curvatura, entonces la suma de
todos los \'{a}ngulos dihedrales alrededor del hueso debe ser $2\pi$. Cuando
en un hueso existe curvatura, la suma de los angulos dihedrales sera
$2\pi-\varepsilon$. Reconocemos entonces que la seccion transversal de este
hueso sera localmente un $\varepsilon-Cono$, y por lo tanto el hueso sera
localmente un $\ \varepsilon-3Cono$. En general en una variedad $\ n$%
-dimensional $M$ , el hueso $T_{n-2}$ tendr\'{a} localmente la geometr\'{i}a
de un $\varepsilon-nCono$.

\bigskip

Ahora bien, un $T_{n-3}$ puede ser en general un extremo de $m$ huesos
$T_{n-2}^{1}T_{n-2}^{2}T_{n-2}^{3}...T_{n-2}^{m}$ . Si estos $m$ se orientan
de manera tal que tengan la direcci\'{o}n positiva saliendo de $T_{n-3}$
entonces se dice que $T_{n-3}$ es una \textit{m-juntura orientada}.

\bigskip En una $m$-juntura orientada existe una relaci\'{o}n entre los
generadores de los huesos $S_{p}$ ($p:1,2,...,m$). Para encontrarla supongamos
una juntura idealizada aislada ( i.e. los huesos se extienden hasta el
infinito). Seleccionamos entonces un orden c\'{i}clico para los huesos. Sea
$A_{p}$ el sector plano que tiene como lados los dos huesos contiguos
$T_{n-2}^{p}$ y $T_{n-2}^{p+1}$. Supondremos que dos de estos sectores no
tienen puntos en com\'{u}n, excepto tal vez los puntos de la frontera (hueso).
De esta manera estos sectores dividen el espacio en dos regiones $M%
\acute{}%
$ y $M%
\acute{}%
\acute{}%
$. Tomamos ahora un punto $P\in M%
\acute{}%
$ y un punto $Q\in M%
\acute{}%
\acute{}%
$ y $m$ trayectorias $t_{p}$ que unen a $P$ con $Q$. Supongamos ahora que la
trayectoria $t_{p}$ intersecta al sector $A_{p}$ una y solo una vez. De esta
manera $t_{p}t_{p-1}^{-1}$ es un aro $a_{p}$ con punto final $P$ y que
encierra al hueso $T_{n-2}^{p}$ una vez. Este aro es un candidato para ser el
generador de $T_{n-2}^{p}$.

El producto de todos los aros formados de esta manera es:%

\begin{equation}
a_{1}a_{2}a_{3}....a_{p}\approx u \label{bianchigeom}%
\end{equation}

\bigskip o en terminos de los generadores de los huesos:%

\begin{equation}
S_{1}S_{2}S_{3}...S_{p}=1 \label{bianchigeom1}%
\end{equation}

\bigskip Como veremos m\'{a}s adelante, la ecuaci\'{o}n \ref{bianchigeom}
correspondera a la aproximacion geom\'{e}trica de la Identidad de Bianchi en
una Variedad Diferenciable Rimanniana.

\bigskip

\section{Variedades Diferencia\label{variedadesdif}les}

\bigskip

Para realizar la transici\'{o}n del Espacio Esqueleto a una variedad
Diferenciable incrementaremos la densidad $\rho$ de Huesos manteniendo el
producto $\rho\varepsilon$ como una funci\'{o}n suave, tal como se
describi\'{o} anteriormente.

Sea $M_{3}$ una variedad tridimensional. Consideremos un haz de Huesos
paralelos en $M_{3}$. Supondremos entonces que la curvatura inducida por los
huesos es peque\~{n}a y por lo tanto $M_{3}$ es aproximadamente Euclideana.Sea
$\mathbf{U}$ un vector unitario paralelo a los huesos. Para medir la curvatura
de la variedad transportamos paralelamente un vector $\mathbf{A}$ alrededor de
un peque\~{n}o aro de area $\Sigma$ y normal unitaria $\widehat{n}$. Es decir
que tenemos $\mathbf{\Sigma}=\Sigma\widehat{n}$. Despues de realizar el TP el
vector $\mathbf{A}$ ha rotado alrededor de $\mathbf{U}$ en un \'{a}ngulo
$\sigma=N\varepsilon$ donde $N$ es el numero de Huesos ecerrados por el aro, y
esta dado por $N=\rho(\mathbf{U\cdot\Sigma)}$. Asi, la diferencia entre el
vector resultante al terminar el TP y el vector inicial ser\'{a}:%

\begin{equation}
\delta\mathbf{A}=\rho\varepsilon(\mathbf{U\cdot\Sigma)(U\wedge A)}
\label{trans1}%
\end{equation}

Ahora bien, de la Relatividad General tenemos que el Transporte Paralelo de un
vector $A_{\mu}$ produce el cambio:

\bigskip%
\begin{subequations}
\begin{equation}
\delta A_{\mu}=R_{\mu\alpha\beta}^{\sigma}\Sigma^{\alpha\beta}A_{\sigma}
\label{trans2}%
\end{equation}

donde $\Sigma^{\alpha\beta}=\epsilon^{\alpha\beta\gamma}\Sigma_{\gamma}$ \ con
$\epsilon^{\alpha\beta\gamma}$ el simbolo de Levi-Civita en tres dimensiones y
$R_{\mu\alpha\beta}^{\sigma}$ es el tensor de Riemann.

Comparando las Ecuaciones \ref{trans1} y \ref{trans2} obtenemos:

\bigskip%

\end{subequations}
\begin{equation}
R_{\mu\sigma\alpha\beta}=\rho\varepsilon U_{\mu\sigma}U_{\alpha\beta}
\label{riemann1}%
\end{equation}

Con $U_{\mu\sigma}=\epsilon_{\mu\sigma\lambda}U^{\lambda}$. Esta
aproximaci\'{o}n tiene todas las propiedades de simetr\'{i}a que posee el
Tensor de Riemann, y ademas cuando existen m\'{a}s haces de huesos, el efecto
es puramente aditivo.

En una variedad $n$-dimensional $M$, un hueso tiene una orientacion
detereminada por un tensor Sesqui-Sim\'{e}trico $U_{\mu\sigma}$ tal que
$U_{\mu\sigma}U^{\mu\sigma}=2$ \ y $U_{\mu\sigma}U_{\alpha\beta}+U_{\mu\alpha
}U_{\beta\sigma}+U_{\mu\beta}U_{\sigma\alpha}=0$. En este caso, si trabajamos
con un haz de huesos paralelos, escogemos las coordenadas $x_{1}$ y $x_{2}$
perpendiculares al haz y las coordenadas $x_{3},x_{4},...,x_{n}$ paralelas al
haz. As\'{i}, tenemos $\ U_{12}=-U_{21}=1$ mientras que las dem\'{a}s
componentes son cero. La Ecuaci\'{o}n \ref{riemann1} a\'{u}n es v\'{a}lida si
hacemos que los indices $\mu,\nu,\alpha,\beta$ tomen los valores
$1,2,3,....,n$.

Por otro lado, independientemente del valor de $n$, el escalar de curvatura
viene dado por:%

\begin{equation}
R=2\rho\varepsilon\label{escalarcurvatura}%
\end{equation}

\bigskip

\subsection{Identidades de Bianchi}

\bigskip

Mostraremos ahora como las Ecuaciones \ \ref{bianchigeom} y \ref{bianchigeom1}
corresponden a las Identidades de Bianchi. Para ello observemos que la
Ecuaci\'{o}n \ref{bianchigeom} nos indica que al recorrer todos los aros
alrededor de los huesos en una $m$-juntura en $M_{3}$ obtenemos el mismo
vector con que comenzamos, es decir que la composici\'{o}n de los $m$ aros
produce la identidad. De esta manera, si las deficiencias de cada hueso de la
juntura $\varepsilon$ son peque\~{n}os, podemos escribir esta relacion para la
$m$-juntura como:%

\begin{equation}
\underset{p=1}{\overset{m}{\sum}}\varepsilon_{p}U_{\mu\nu}^{p}=0
\label{bianchigeo}%
\end{equation}

o equivalentemente

\bigskip%

\begin{equation}
\underset{p=1}{\overset{m}{\sum}}\varepsilon_{p}\mathbf{U}^{p}=0
\end{equation}

\bigskip

Ahora bien, supongamos una distribuci\'{o}n de densidad $\varrho$ de
$m$-junturas identicas. Es decir que tendremos $m$ haces de huesos paralelos.
Supongamos entonces que el $p$-esimo haz de huesos tiene una deficiencia
$\varepsilon_{p}$, una direcci\'{o}n dada por el vector\textbf{\ }%
$\mathbf{U}^{p}$ y una densidad $\rho_{p}$. Es claro que $\rho_{p}$ no es
constante, ya que cada juntura hace un decaimiento o un crecimiento de esta
densidad con una razon que depende de la densidad de junturas $\varrho$.

Por definici\'{o}n tenemos que $\rho_{p}$ es la densidad de flujo de huesos
que salen de la juntura a traves de una superficie $\Sigma$ ortogonal a
$\mathbf{U}^{p}$. Supongamos ahora que la posici\'{o}n de la superficie
$\Sigma$ esta determinada por una abscisa $s$ que se mide sobre la
direcci\'{o}n del vector $\mathbf{U}_{p}$. Sea entonces $C$ un cilindro de
alura $ds$ y base $\Sigma(s)$. \ As\'{i}, cada juntura dentro de $C$
incrementa en una unidad el flujo de $\rho_{p}$ huesos a traves de la
superficie $\Sigma(s+ds)$. Adem\'{a}s dentro del cilindro $C$ existiran
$\Sigma\cdot\varrho\cdot ds$ junturas. De esta manera tendremos:%

\begin{equation}
\Sigma\rho_{p}(s+ds)-\Sigma\rho_{p}(s)=\Sigma\varrho(s)ds
\end{equation}

\bigskip

es decir:

\bigskip%

\begin{equation}
\frac{d\rho_{p}}{ds}=\varrho(s)
\end{equation}

\bigskip

En general, esta expresi\'{o}n ser\'{a}:

\bigskip%

\begin{equation}
\mathbf{U}^{p,\mu}\cdot\mathbf{\partial}_{\mu}\rho_{p}=\varrho
\label{bianchiaux2}%
\end{equation}

\bigskip

Ahora bien, el Tensor de Riemann vendra dado por la expresi\'{o}n
\ref{riemann1} pero debemos sumar sobre todas las junturas, es decir:

\bigskip%

\begin{equation}
R_{\alpha\beta\lambda\delta}=\underset{p}{\sum}\varepsilon_{p}\rho
_{p}U_{\alpha\beta}^{p}U_{\lambda\delta}^{p} \label{riemann2}%
\end{equation}

\bigskip

Ya que hemos supuesto que la curvatura inducida por los huesos es peque\~{n}a,
podemos aproximar tambien las derivadas covariantes con derivadas normales, y
de esta manera las identidades de Bianchi ser\'{a}n:

\bigskip%

\begin{equation}
B_{\alpha\beta\lambda\delta\gamma}=\frac{\partial}{\partial x^{\gamma}%
}R_{\alpha\beta\lambda\delta}+\frac{\partial}{\partial x^{\lambda}}%
R_{\alpha\beta\delta\gamma}+\frac{\partial}{\partial x^{\delta}}R_{\alpha
\beta\gamma\lambda}=0 \label{bianchi1}%
\end{equation}

\bigskip

Reemplazando la Ecuaci\'{o}n \ref{riemann2} en la Ecuacion \ref{bianchi1} y
tratando $\varepsilon_{p}$ y $U^{p}$ como constantes tenemos:

\bigskip%

\begin{equation}
B_{\alpha\beta\lambda\delta\gamma}=\underset{p}{\sum}\varepsilon_{p}%
U_{\alpha\beta}^{p}U_{\lambda\delta}^{p}\frac{\partial\rho_{p}}{\partial
x^{\gamma}}+\underset{p}{\sum}\varepsilon_{p}U_{\alpha\beta}^{p}%
U_{\delta\gamma}^{p}\frac{\partial\rho_{p}}{\partial x^{\lambda}}+\underset
{p}{\sum}\varepsilon_{p}U_{\alpha\beta}^{p}U_{\gamma\lambda}^{p}\frac
{\partial\rho_{p}}{\partial x^{\delta}}=0
\end{equation}

\bigskip es decir:

\bigskip%

\begin{equation}
B_{\alpha\beta\lambda\delta\gamma}=\underset{p}{\sum}\varepsilon_{p}%
U_{\alpha\beta}^{p}\left[  U_{\lambda\delta}^{p}\frac{\partial}{\partial
x^{\gamma}}+U_{\delta\gamma}^{p}\frac{\partial}{\partial x^{\lambda}%
}+U_{\gamma\lambda}^{p}\frac{\partial}{\partial x^{\delta}}\right]  \rho_{p}=0
\label{bianchiaux1}%
\end{equation}

\bigskip

Es inmediato verificar que se tiene la identidad entre operadores:

\bigskip%

\begin{equation}
U_{\lambda\delta}^{{}}\frac{\partial}{\partial x^{\gamma}}+U_{\delta\gamma
}^{{}}\frac{\partial}{\partial x^{\lambda}}+U_{\gamma\lambda}^{{}}%
\frac{\partial}{\partial x^{\delta}}=\varepsilon_{\gamma\lambda\delta}%
U^{v}\frac{\partial}{\partial x^{v}}%
\end{equation}%

\begin{equation}
U_{\lambda\delta}^{{}}\frac{\partial}{\partial x^{\gamma}}+U_{\delta\gamma
}^{{}}\frac{\partial}{\partial x^{\lambda}}+U_{\gamma\lambda}^{{}}%
\frac{\partial}{\partial x^{\delta}}=\varepsilon_{\gamma\lambda\delta}%
U^{v}\partial_{v}%
\end{equation}

\bigskip

Reemplazando esta identidad en \ref{bianchiaux1} tenemos:

\bigskip%

\begin{equation}
B_{\alpha\beta\lambda\delta\gamma}=\underset{p}{\sum}\varepsilon_{p}%
U_{\alpha\beta}^{p}\left[  \varepsilon_{\gamma\lambda\delta}U^{p,v}%
\partial_{v}\right]  \rho_{p}=0
\end{equation}%

\begin{equation}
B_{\alpha\beta\lambda\delta\gamma}=\varepsilon_{\gamma\lambda\delta}%
\underset{p}{\sum}\varepsilon_{p}U_{\alpha\beta}^{p}U^{p,v}\partial_{v}%
\rho_{p}=0
\end{equation}

\bigskip

Haciendo uso de \ref{bianchiaux2} tenemos:

\bigskip%

\begin{equation}
B_{\alpha\beta\lambda\delta\gamma}=\varrho\varepsilon_{\gamma\lambda\delta
}\underset{p}{\sum}\varepsilon_{p}U_{\alpha\beta}^{p}=0
\end{equation}

\bigskip

que coincide exactamente con \ref{bianchigeo} y por lo tanto nos muestra la
relaci\'{o}n de \ref{bianchigeom} y de \ref{bianchigeom1} con las
identidadesde Bianchi.

\bigskip

\section{\bigskip Variedades Lorentizianas}

Cuando la signatura del tensor m\'{e}trico $\mathbf{g}$ de una variedad
$n$-dimensional $M_{n}$ , es $sig(\mathbf{g})=2-n$ la variedad se denomina
Lorentziana. En este caso, la m\'{e}trica de la variedad no es siempre
positiva. En este caso, la representacion del grupo fundamental, es decir las
matrices ortogonales $S$ deben ser reemplazadas por matrices de Lorentz.
Correspondientemente, en el Espacio Esqueleto, los \'{a}ngulos de deficit
pueden ser imaginarios.

\bigskip

Nos restringiremos ahora al caso 4-dimensional. En esta variedad, los huesos
son tri\'{a}ngulos, y la m\'{e}trica tiene una coordenada temporal. De esta
manera tenemos tres clases de huesos:

\begin{enumerate}
\item \textit{\ Como-de-Espacio:} Cuando cualquier vector en el hueso es como
de espacio. El \'{a}ngulo de deficit $\varepsilon$ es imaginario.

\item \textit{Nulo}: Cada vector en el hueso es una combinaci\'{o}n lineal de
un vector nulo y un vector como de espacio ortogonal a \'{e}l. En este caso
tenemos $\varepsilon=0$ .

\item \textit{Como-de-Tiempo}: Existen vectores como-de-tiempo y
como-de-espacio en el hueso. $\varepsilon$ es real.
\end{enumerate}

\bigskip

Ahora definiremos el area $L$ de un hueso. Para ello, notemos que en $M_{3}$
el \'{a}rea de un tri\'{a}ngulo que tiene como lados los vectores $\mathbf{B}$
y $\mathbf{C}$ viene descrita por el vector:

\bigskip%

\begin{equation}
2\mathbf{A=B\times C}%
\end{equation}

\bigskip

que es perpendicular a $\mathbf{B}$ y a $\mathbf{C}$. La magnitud del \'{a}rea sera:

\bigskip%

\begin{equation}
4A^{2}=B^{2}C^{2}\sin^{2}\theta
\end{equation}

donde $\theta$ es el \'{a}ngulo formado por los vectores $\mathbf{B}$ y
$\mathbf{C}$. Tenemos entonces:

\bigskip%

\begin{equation}
4A^{2}=B^{2}C^{2}(1-\cos^{2}\theta)=B^{2}C^{2}-B^{2}C^{2}\cos^{2}\theta
\end{equation}%

\begin{equation}
4A^{2}=B^{2}C^{2}-\left(  \mathbf{B\cdot C}\right)  ^{2}%
\end{equation}

\bigskip

En el espacio 4-dimensional, definiremos entonces el \'{a}rea $L$ como la
generalizaci\'{o}n:

\bigskip%

\begin{equation}
4L^{2}=\left(  B_{\mu}C^{\mu}\right)  ^{2}-(B^{\mu}B_{\mu})(C^{\mu}C_{\mu})
\end{equation}

\bigskip

Se puede comprobar entonces que al igual que en el caso 3-dimensional, el
vector $2\mathbf{L}$ es dual al bivector construido de $\mathbf{B}$ a
$\mathbf{C}$. Es decir que si $\mathbf{B}$ va en direcci\'{o}n $x^{0}$ y
$\mathbf{C}$ va en direccion $x^{3},$ entonces $\mathbf{L}$ es un bivector en
el plano $(x^{1},x^{2})$. Asi, $L$ es una cantidad real siempre que
$\mathbf{B}$ y $\mathbf{C}$ no sean ambos como-de-espacio. Por lo tanto
trabajaremos \'{u}nicamente con huesos nulos y como-de-tiempo. Adem\'{a}s es
claro que si el hueso es nulo, tenemos $L=0.$

\bigskip

\chapter{\bigskip C\'{a}lculo de Regge y Relatividad General}

\section{Ecuaciones de Campo de Einstein en el Vacio}

\bigskip

Utilizando todas las anteriores definiciones y el esquema anterior del
C\'{a}lculo en Espacios Esqueleto, deduciremos ahora el an\'{a}logo
discretizado de las Ecuaciones de Campo de Einstein para el Vacio.

La manera m\'{a}s directa de llegar a la esqueletizaci\'{o}n de un espacio de
Eintein y de llegar a las ecuaciones de campo es partiendo de un principio
variacional. Para ello tomamos la Acci\'{o}n de Einstein-Hilbert:

\bigskip%

\begin{equation}
I=\frac{1}{16\pi}\int R\sqrt{-g}d^{4}x
\end{equation}

\bigskip

Las ecuaciones de de Campo de Einstein corresponderan entonces a la
condici\'{o}n de que $I$ sea un extremo. Es decir $\delta I=0.$ Esta
condici\'{o}n es aplicable solamente cuando consideramos un espacio libre de
materia y de campo electromagn\'{e}tico, es decir que obtendremos las
ecuaciones de campo para el espacio vacio.

En el espacio esqueleto, la variaci\'{o}n de esta acci\'{o}n debe realizarse
sobre las longitudes de los simplex $T_{1}$, ya que como sabemos estas
longitudes contienen la misma informaci\'{o}n que el tensor m\'{e}trico.

\bigskip

Ahora bien, para obtener una expresi\'{o}n de $I$ en t\'{e}rminos de las
longitudes, observemos primero que ni $T_{4}$ ni $T_{3}$ contribuyen a la
integral, ya que $R$ es simplemente una funci\'{o}n distribuci\'{o}n con
soporte $w$ tal como vimos anteriormente.\ Por lo tanto $I$ debe ser una
funci\'{o}n de los huesos $T_{2}$. Si $k$ es el n\'{u}mero de huesos en el
esqueleto tenemos $I=\underset{k}{\sum}F(T_{2}^{k})$ . Ahora bien, el \'{a}rea
del hueso $L$ es una funci\'{o}n aditiva, mientras que la funci\'{o}n $F$ es
la misma para todos los huesos. Supondremos entonces que $F$ es proporcional
al \'{a}rea del hueso, i.e. $F(T_{2}^{k})=L_{k}f(\varepsilon_{k})$.

\bigskip

Por otro lado, un hueso $T_{2}^{k}$ puede ser expresado como la
superposici\'{o}n de dos huesos $T_{2}^{k}%
\acute{}%
$ y $T_{2}^{k}%
\acute{}%
\acute{}%
$ con la misma forma y \'{a}rea y tales que $\varepsilon_{k}=\varepsilon_{k}%
\acute{}%
+\varepsilon_{k}%
\acute{}%
\acute{}%
$ . Se sigue entonces que $f(\varepsilon_{k})=f(\varepsilon_{k}%
\acute{}%
)+f(\varepsilon_{k}%
\acute{}%
\acute{}%
)$. \ Esta condici\'{o}n implica que $f(\varepsilon)=C\varepsilon$ donde $C$
es una constante. As\'{i} tenemos que la acci\'{o}n se puede expresar como:

\bigskip%

\begin{equation}
I=C\underset{k}{\sum}L_{k}\varepsilon_{k} \label{accionesqueleto}%
\end{equation}

\bigskip

Utilizando la expresi\'{o}n para el escalar de curvatura dado por la
ecuaci\'{o}n \ref{escalarcurvatura} y en los l\'{i}mites de una variedad
aproximadamente plana tenemos que la acci\'{o}n viene dada por:

\bigskip%

\begin{equation}
I=\frac{1}{8\pi}\int\varepsilon\varrho d^{4}x \label{accionvariedad}%
\end{equation}

\bigskip

Comparando las ecuaciones \ref{accionesqueleto} y \ref{accionvariedad} tenemos
$C=\frac{1}{8\pi}$, con lo que obtenemos finalmente la Acci\'{o}n de Regge:

\bigskip%

\begin{equation}
I=\frac{1}{8\pi}\underset{k}{\sum}L_{k}\varepsilon_{k} \label{accionregge}%
\end{equation}

\bigskip

En una variedad $n$-dimensional $M_{n\text{ }}$ tendremos la generalizaci\'{o}n:

\bigskip%
\begin{equation}
I=\frac{1}{8\pi}\underset{k}{\sum}L_{k}^{n}\varepsilon_{k}%
\end{equation}

\bigskip donde $L_{k}^{n}$ es la $(n-2)-$dimensional medida del hueso
$T_{n-2}$.

\bigskip

La siguiente tarea ser\'{a} encontrar la variaci\'{o}n de la Acci\'{o}n de
Regge en t\'{e}rminos de las longitudes $l_{p}$\bigskip\ de los simplex
$T_{1}^{p}$ constituyentes de los huesos. Esto es:

\bigskip

\bigskip%
\begin{equation}
\delta I=\frac{1}{8\pi}\delta\underset{k}{\sum}L_{k}\varepsilon_{k}=0
\label{variacion1}%
\end{equation}

\bigskip

La dependencia de el \'{a}rea del hueso en esta longitud es clara, pero la
dependencia del \'{a}ngulo de d\'{e}ficit $\varepsilon$ es bastante
complicada. Sin embargo, siguiendo a T. Regge$^{\cite{regge}}$ se puede
comprobar (ver Ap\'{e}ndice 1) que la variaci\'{o}n puede realizarse como si
los $\varepsilon_{n}$ fuesen constantes, es decir que la ecuaci\'{o}n
\ref{variacion1} es simplemente:

\bigskip%

\begin{equation}
\delta I=\frac{1}{8\pi}\underset{k}{\sum}\varepsilon_{k}\delta L_{k}=0
\label{variacion2}%
\end{equation}

\bigskip

En el caso de $M_{3}$ los huesos seran los v\'{e}rtices y por lo tanto $\delta
L_{k}=\delta l_{k}$, y por lo tanto la Ecuaci\'{o}n \ref{variacion2} se reduce
simplemente a la condici\'{o}n $\varepsilon_{k}=0$. Esto quiere decir que
todos los espacios esqueletos de Einstein de dimensi\'{o}n 3 son planos.

\bigskip

Ahora bien, en una variedad 4-dimensional, y de acuerdo con la
trigonometr\'{i}a elemental, el cambio en el \'{a}rea de un tri\'{a}ngulo en
funci\'{o}n de la variaci\'{o}n en la longitud de sus lados ser\'{a} (ver
Ap\'{e}ndice 2):

\bigskip%

\begin{equation}
\delta L_{k}=\frac{1}{2}\underset{p}{\sum}l_{p}\cot\left(  \theta_{pk}\right)
\delta l_{p}%
\end{equation}

Donde $\theta_{pk}$ es el \'{a}ngulo opuesto al lado $l_{p}$ en el
tri\'{a}ngulo $T_{2}^{k}$. De esta manera reemplazando este resultado en la
Ecuaci\~{n}on \ref{variacion2} obtenemos:

\bigskip%

\begin{equation}
\delta I=\frac{1}{16\pi}\underset{k}{\sum}\underset{p}{\sum}\varepsilon
_{k}l_{p}\cot\left(  \theta_{pk}\right)  \delta l_{p}=0
\end{equation}

\bigskip

es decir:

\bigskip%

\begin{equation}
\underset{k}{\sum}\varepsilon_{k}\cot\left(  \theta_{pk}\right)  =0\text{
\ \ \ \ \ \ \ \ con \ \ }p=1,2,.... \label{einsteinregge}%
\end{equation}

\bigskip

La ecuaci\'{o}n \ref{einsteinregge} corresponde a la versi\'{o}n discretizada
de las Ecuaciones de Campo de Einstein en el vacio seg\'{u}n el m\'{e}todo de
Regge. Se puede observar que se tiene una ecuaci\'{o}n por cada longitud de un
v\'{e}rtice dentro de la regi\'{o}n de espacio-tiempo que se analiza.

\bigskip

\section{\bigskip Desviaci\'{o}n de la Geodesicas en el Modelo de Regge}

Una de las preguntas abiertas en la formulaci\'{o}n de Regge es la
interacci\'{o}n entre el campo Gravitacional y materia. Se han logrado
realizar la inclusi\'{o}n de algunos t\'{e}rminos que representan materia en
esta formulaci\'{o}n, pero para ello se deben hacer algunas restricciones
acerca de la clase de materia que se acopla al campo gravitacional y se deben
imponer ciertas simetr\'{i}as al campo.

Otra manera de incorporar la materia en la formulaci\'{o}n de Regge es la de
identificar primero los par\'{a}metros que identifican a la materia tanto en
la formulaci\'{o}n continua de la Relatividad General como en el c\'{a}lculo
de Regge, y de esta manera simplemente se insertar\'{i}an los t\'{e}rminos
apropiados en las ecuaciones discretas. Sin embargo esta identificaci\'{o}n no
es trivial y usualmente conduce a divergencias.

Una manera de comenzar con esta identificaci\'{o}n es analizar inicialmente
una de las manifestaciones mas elementales de la influencia de la gravedad en
la materia: la Desviaci\'{o}n de Geod\'{e}sicas. Esta desviaci\'{o}n nos
describe la aceleraci\'{o}n relativa de part\'{i}culas de prueba libres en un
espacio-tiempo curvo. Primero analizaremos el caso en la formulaci\'{o}n
continua siguiendo a Synge$^{\cite{synge}}$.

\bigskip

\subsection{Desviaci\'{o}n de Geodesicas en Relatividad General}

\bigskip

Sea $C(v)$ un conjunto de curvas con ecuaci\'{o}n $x^{\mu}=x^{\mu}(u,v)$ , de
tal manera que $v$ es constante a lo largo de cada curva, y parametrizada por
$u$. Este conjunto de curvas forman una superficie 2-dimensional.

\bigskip%

\begin{figure}
[h]
\begin{center}
\includegraphics[
natheight=177.000000pt,
natwidth=255.375000pt,
height=127.125pt,
width=182.8125pt
]%
{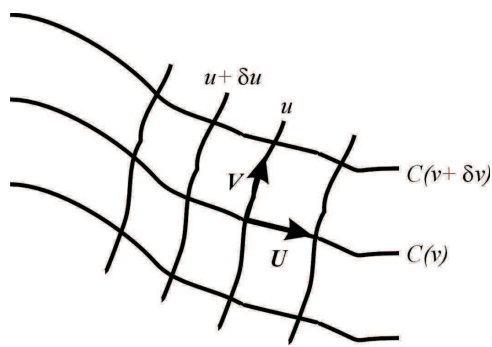}%
\caption{Familia de Geodesicas $C(v)$ parametrizadas por $u$. Los vectores
$\mathbf{U}$ y $\mathbf{V}$ son tangentes a las curvas con $v$ y $u$ constante
respectivamente.}%
\label{desviacion1}%
\end{center}
\end{figure}

\bigskip

Los vectores $\ \ \ \ U^{\mu}=\partial_{u}x^{\mu}=\frac{\partial x^{\mu}%
}{\partial u}$ \ \ y \ $V^{\mu}=\partial_{v}x^{\mu}=\frac{\partial x^{\mu}%
}{\partial v}\ $\ satisfacen:%

\begin{equation}
\frac{\delta U^{\mu}}{\delta v}=\frac{\delta V^{\mu}}{\delta u}
\label{geodes1}%
\end{equation}

donde $\frac{\delta}{\delta u}$ y $\frac{\delta}{\delta v}$ son las derivadas
covariantes a lo largo de las curvas con $v$ y $u$ constante respectivamente.

\bigskip

Consideremos dos curvas adyacentes $C(v)$ \ y $C(v+\delta v)$. \ De esta
manera, el vector $V^{\mu}$ representara la separaci\'{o}n entre las dos
curvas. Para encontrar como se desvia la curva $C(v+\delta v)$ de la curva
$C(v)$ calculamos la segunda derivada covariante del vector de separaci\'{o}n,
a lo largo de la curva $C(v)$, parametrizada por $u$. Esto es:

\bigskip%

\begin{equation}
\frac{\delta^{2}V^{\mu}}{\delta u^{2}}=\frac{\delta}{\delta u}\frac{\delta
V^{\mu}}{\delta u}%
\end{equation}

\bigskip Utilizando \ref{geodes1} tenemos:%

\begin{equation}
\frac{\delta^{2}V^{\mu}}{\delta u^{2}}=\frac{\delta}{\delta u}\frac{\delta
U^{\mu}}{\delta v} \label{geodes2}%
\end{equation}

\bigskip

Ahora bien, sabemos que la diferenciaci\'{o}n covariante en una 2-superficie
satisface la regla de conmutaci\'{o}n:

\bigskip%

\begin{equation}
\frac{\delta^{2}T^{\mu}}{\delta u\delta v}-\frac{\delta^{2}T^{\mu}}{\delta
v\delta u}=R_{\alpha\beta\gamma}^{\mu}T^{\alpha}U^{\beta}V^{\gamma}%
\end{equation}

\bigskip

As\'{i} tenemos que la ecuaci\'{o}n \ref{geodes2} ser\'{a}:

\bigskip%

\begin{equation}
\frac{\delta^{2}V^{\mu}}{\delta u^{2}}=\frac{\delta}{\delta v}\frac{\delta
U^{\mu}}{\delta u}+R_{\alpha\beta\gamma}^{\mu}U^{\alpha}U^{\beta}V^{\gamma}%
\end{equation}

\bigskip

donde $R_{\alpha\beta\gamma}^{\mu}$ es el tensor de Riemann.

Si tomamos las curvas $C(v)$ como geod\'{e}sicas tenemos la condici\'{o}n:%

\begin{equation}
\frac{\delta U^{\mu}}{\delta u}=0
\end{equation}

As\'{i}, tenemos que la ecuaci\'{o}n de la desviaci\'{o}n de las geodesicas
ser\'{a}:

\bigskip%

\begin{align}
\frac{\delta^{2}V^{\mu}}{\delta u^{2}}  &  =R_{\alpha\beta\gamma}^{\mu
}U^{\alpha}U^{\beta}V^{\gamma}\\
\frac{\delta^{2}V^{\mu}}{\delta u^{2}}-R_{\alpha\beta\gamma}^{\mu}U^{\alpha
}U^{\beta}V^{\gamma}  &  =0
\end{align}

\bigskip Adem\'{a}s, utilizando las simetr\'{i}as del Tensor de Riemann tenemos:%

\begin{equation}
\frac{\delta^{2}V^{\mu}}{\delta u^{2}}+R_{\alpha\beta\gamma}^{\mu}U^{\alpha
}V^{\beta}U^{\gamma}=0 \label{desviaciongeodesicacont}%
\end{equation}

Ahora bien,si en lugar del par\'{a}metro $u$ utilizamos la longitud de arco
$s$, y las lineas parametrizadas con $v$ las tomamos ortogonales a las $C(v)$;
la ecuaci\'{o}n de desviaci\'{o}n de Geod\'{e}sicas toma la forma general:

\bigskip%

\begin{equation}
\frac{\delta^{2}\mathbf{V}}{\delta s^{2}}+\mathbf{R(U,V,U)}=0
\label{desviaciongeneral}%
\end{equation}

Como caso especial, en una variedad 2-dimensional el tensor de Riemann
tendr\'{a} \'{u}nicamente una componente diferente de cero, que tomaremos como
$R_{212}^{1}$.

\bigskip

\subsection{Desviaci\'{o}n de las Geod\'{e}sicas en el modelo discretizado:}

\bigskip

Para observar el efecto de la desviaci\'{o}n de las geod\'{e}sicas en el caso
discreto seguiremos el trabajo de S. Chakrabarti, et. al.$^{\cite{geodesic}}$.
En este caso ser\'{a} suficiente considerar un espacio-tiempo 2-dimensional.
De esta manera la curvatura estar\'{a} concentrada en los v\'{e}rtices de la
red simplicial.

\bigskip%

\begin{figure}
[h]
\begin{center}
\includegraphics[
natheight=178.437500pt,
natwidth=407.812500pt,
height=111.8125pt,
width=253.9375pt
]%
{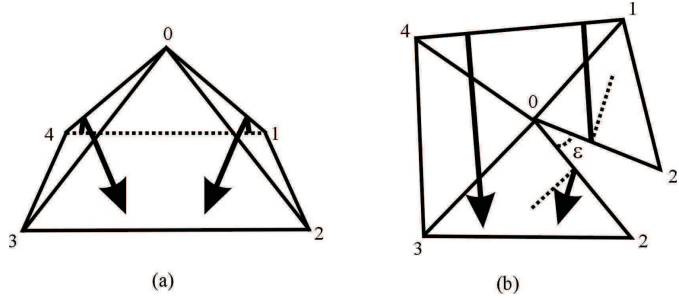}%
\caption{(a) Dos geodesicas originalmente paralelas, convergen cuando entran
al triangulo [023] debido al angulo de deficit $\varepsilon$ en el vertice
[0]. (b) Al realizar un corte en la arista [02] y colocando el simplex en un
plano se observa el angulo de deficit $\varepsilon$. Las geodesicas son ahora
rectas, y no cambian de direccion al pasar de un triangulo al otro. La
desviacion solo se hace visible en el triangulo [023].}%
\label{desviacionsimplex}%
\end{center}
\end{figure}

\bigskip

En la figura \ref{desviacionsimplex}(a) vemos como dos geod\'{e}sicas que son
paralelas al entrar en el tri\'{a}ngulo [014] sufren una desviaci\'{o}n debido
al \'{a}ngulo de d\'{e}ficit $\varepsilon$ del v\'{e}rtice [0]. Para obtener
una mejor visi\'{o}n de lo que sucede en realidad realizamos un corte en una
de las aristas del $3-$simplex \ [01234] y colocamos la figura resultante en
un plano (Figura \ref{desviacionsimplex}(b)). As\'{i}, las geod\'{e}sicas se
convierten en simples rectas. Sin embargo, a\'{u}n cuando inicialmente son
paralelas, al llegar al tri\'{a}ngulo [023] convergen debido al \'{a}ngulo de
d\'{e}ficit $\varepsilon$.

Observamos entonces que la pasar alrededor de un v\'{e}rtice, las
geod\'{e}sicas sufren una desviaci\'{o}n \ que se manifiesta en una cambio en
el \'{a}ngulo de convergencia de las geod\'{e}sicas $\Delta\alpha$ , y este
cambio corresponde al \'{a}ngulo de d\'{e}ficit asociado con el v\'{e}rtice, i.e.

\bigskip%

\begin{equation}
\Delta\alpha=\varepsilon
\end{equation}

\bigskip

Es decir que al igual que en el caso continuo, la curvatura enfoca o dispersa
las geod\'{e}sicas. Sin embargo, en el caso de un \'{u}nico v\'{e}rtice, el
cambio en el \'{a}ngulo de convergencia de las geod\'{e}sicas no puede ser
continuo. As\'{i}, no es posible encontrar una ecuaci\'{o}n de desviaci\'{o}n
de geod\'{e}sicas que dependa de la segunda derivada del vector de
separaci\'{o}n, tal como en el caso continuo.

\subsubsection{Desviaci\'{o}n Producida por una Distribuci\'{o}n de
V\'{e}rtices (Caso 2-dimensional)}

\bigskip

Para lograr encontrar el an\'{a}logo de la Ecuaci\'{o}n de Desviaci\'{o}n de
Geod\'{e}sicas (Ec. \ref{desviaciongeodesicacont} ) en la formulaci\'{o}n de
Regge, estudiaremos el caso de un espacio 2-dimensional.

\bigskip

Consideramos una region de una red Simplicial, en la cual existen $N$
v\'{e}rtices con \'{a}ngulos de d\'{e}ficit $\{\varepsilon_{i}\}_{i=1}^{N}$ .
Asumimos adem\'{a}s que cada \'{a}ngulo de d\'{e}ficit $\varepsilon_{i}$ es
peque\~{n}o, y que el \'{a}rea que contiene los v\'{e}rtices, $\Delta A,$ es
tambi\'{e}n peque\~{n}a. Sin embargo, el \'{a}ngulo total de d\'{e}ficit por
unidad de area

\bigskip%

\begin{equation}
\Theta=\frac{1}{\Delta A}\underset{i=1}{\overset{N}{\sum}}\varepsilon
_{i}=n\overline{\varepsilon} \label{deficittotal}%
\end{equation}

es finito. Adem\'{a}s hemos definido $n$ como la densidad de v\'{e}rtices por
unidad de \'{a}rea y $\overline{\varepsilon}$ como el \'{a}ngulo de
d\'{e}ficit promedio por v\'{e}rtice:

\bigskip%

\begin{equation}
\overline{\varepsilon}=\frac{1}{N}\underset{i=1}{\overset{N}{\sum}}%
\varepsilon_{i}%
\end{equation}

Consideremos ahora dos geod\'{e}sicas $C_{1}$ y $C_{2}$ que pasan rodeando el
conjunto de $N$ v\'{e}rtices. El cambio en el \'{a}ngulo de convergencia de
las dos geod\'{e}sicas estar\'{a} determinado por el \'{a}ngulo total de
d\'{e}ficit asociado con el \'{a}rea considerada:

\bigskip%

\begin{equation}
\Delta\alpha=\underset{i=1}{\overset{N}{\sum}}\varepsilon_{i}
\label{angulocon}%
\end{equation}

\bigskip

Asumiremos ahora que cada geod\'{e}sica esta parametrizada por la longitud de
arco $s$ y que estas geod\'{e}sicas estan definidas adem\'{a}s por un
par\'{a}metro $v$ tal que las lineas $s=const.$ son ortogonales a las
geod\'{e}sicas. Aproximamos ahora esta situaci\'{o}n a una variedad
2-dimensional con coordenadas $(x^{1},x^{2})$, tales que las lineas con
coordenada $x^{1}=const.$ corresponden a las curvas con $v=const.$ mientras
que las lineas con $x^{2}=const.$ corresponden a las geod\'{e}sicas.

\bigskip

\bigskip%
\begin{figure}
[h]
\begin{center}
\includegraphics[
natheight=133.937500pt,
natwidth=423.750000pt,
height=98.875pt,
width=309.9375pt
]%
{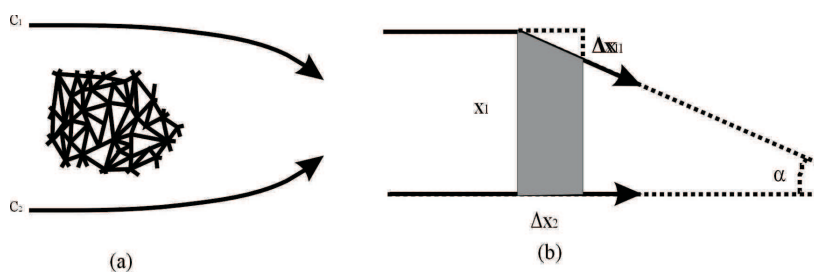}%
\caption{(a) Las geodesicas $C_{1}$ y $C_{2}$ son defelctadas por $N$
vertices. \ (b) Representacion de la desviacion de geodesicas producidas por
la distribucion de $N$ vertices.}%
\label{desviacioncomplx}%
\end{center}
\end{figure}

\bigskip

De esta manera, el \'{a}rea que contiene los $N$ v\'{e}rtices esta dada por
$\Delta A=x^{1}\Delta x^{2}m.$(Figura \ \ref{desviacioncomplx}\ (b)).

Las componentes del Vector de Separaci\'{o}n $\mathbf{V}$ y del vector
velocidad de la geod\'{e}sica $\mathbf{U}$ ser\'{a}n:

\bigskip%

\begin{align}
\mathbf{V}  &  =(x^{1},0)\\
\mathbf{U}  &  =\left(  0,\frac{\Delta x^{2}}{\Delta s}\right)  =(0,1)
\end{align}

\bigskip

Adem\'{a}s, de la Figura \ref{desviacioncomplx}(b)\ \ podemos observar que el
\'{a}ngulo de convergencia se puede aproximar por:

\bigskip%

\begin{equation}
\alpha\approx\tan\alpha=\frac{\Delta x^{1}}{\Delta x^{2}} \label{angulocongru}%
\end{equation}

\bigskip

De esta manera, utilizando las expresiones \ref{angulocongru} y
\ref{deficittotal} podemos escribir la ecuaci\'{o}n \ref{angulocon} como:

\bigskip%

\begin{align}
\Delta\left(  \frac{\Delta x^{1}}{\Delta x^{2}}\right)   &  =\underset
{i=1}{\overset{N}{\sum}}\varepsilon_{i}=n\overline{\varepsilon}\Delta A\\
\Delta\left(  \frac{\Delta x^{1}}{\Delta x^{2}}\right)   &  =n\overline
{\varepsilon}x^{1}\Delta x^{2}%
\end{align}

\bigskip

Ya que $\Delta x^{2}=\Delta s$ y como la norma del vector de separacion es
$V=x^{1}$ tenemos:

\bigskip%

\begin{align}
\Delta\left(  \frac{\Delta V}{\Delta s}\right)  +n\overline{\varepsilon
}V\Delta s  &  =0\\
\frac{\Delta}{\Delta s}\left(  \frac{\Delta V}{\Delta s}\right)
+n\overline{\varepsilon}V  &  =0
\end{align}

\bigskip

Llevando esta expresi\'{o}n al caso infinitesimal, y ya que el vector
separaci\'{o}n puede expresarse como $V=(V^{1},0)$ obtenemos:

\bigskip%

\begin{equation}
\frac{\delta^{2}V^{1}}{\delta s^{2}}+n\overline{\varepsilon}V^{1}=0
\end{equation}

\bigskip

Ahora bien, como dijimos en la secci\'{o}n anterior, en el caso 2-dimensional
el tensor de Riemann tiene una sola componente diferente de cero, que
tomaremos \ como $R_{212}^{1}$ , y de esta manera la Ecuaci\'{o}n
\ \ref{desviaciongeneral} puede escribirse como:

\bigskip%

\begin{equation}
\frac{\delta^{2}V^{1}}{\delta s^{2}}+R_{212}^{1}V^{1}=0
\end{equation}

Comparando estas dos ultimas ecuaciones obtenemos la identificaci\'{o}n:

\bigskip%

\begin{equation}
R_{212}^{1}=n\overline{\varepsilon}%
\end{equation}

es decir que la \'{u}nica componente del Tensor de Riemann en el caso
2-dimensional se puede interprestar como el \'{a}ngulo de d\'{e}ficit promedio
por unidad de \'{a}rea.

\bigskip

\subsubsection{Generalizaci\'{o}n a 4 dimensiones}

\bigskip

Para generalizar los anteriores resultados a una variedad 4-dimensional,
simplemente debemos notar que en este caso la curvatura esta concentrada en
$2-$simplex, es decir tri\'{a}ngulos. Adem\'{a}s, para una variedad
4-dimensional el tensor de Riemann tiene 20 componentes independientes.

\bigskip

Si tenemos una regi\'{o}n infinitesiomal con $N$ tri\'{a}ngulos, una
geod\'{e}sica que se encuentre dentro de un plano ortogonal a uno de esos
tri\'{a}ngulos sufrir\'{a} una desviaci\'{o}n igual a la magnitud del
\'{a}ngulo de d\'{e}ficit correspondiente al tri\'{a}ngulo. Es decir
exactamente igual al caso 2-dimensional.

Ahora bien, en general la geod\'{e}sica no va a encontrarse en el plano
ortogonal a todos los triangulos, y adem\'{a}s no todos los triangulos tienen
la misma orientaci\'{o}n. Si definimos el \'{a}ngulo $\theta$ \ como el
formado entre la direcci\'{o}n del vector tangente a la geod\'{e}sica y el
vector normal a uno de los triangulos, tenemos que el vector tangente
sufrir\'{a} una rotaci\'{o}n en un \'{a}ngulo $\varepsilon\cos\theta$.

De esta manera es inmediato que para hacer la generalizaci\'{o}n a 4
dimensiones solamente debemos reemplazar $n$ por el n\'{u}mero de triangulos
por unidad de volumen y cambiar el \'{a}ngulo d\'{e}ficit promedio
$\overline{\varepsilon}$ por el \'{a}ngulo de d\'{e}ficit efectivo que afecta
una geod\'{e}sica con vector tangente $\mathbf{U}$ definido como:

\bigskip%

\begin{equation}
\varepsilon_{U}=\underset{i=1}{\overset{N}{\sum}}\varepsilon_{i}\cos\theta_{i}%
\end{equation}

donde $\theta_{i}$ es el \'{a}ngulo entre el vector tangente $U$ y el plano
ortogonal al tri\'{a}ngulo con \'{a}ngulo de d\'{e}ficit $\varepsilon_{i}$.

\bigskip

De esta manera recuperamos la definici\'{o}n del tensor de Riemann obtenida en
la Ecuaci\'{o}n \ref{riemann1}, es decir que en la formulaci\'{o}n de Regge
esta representado por el \'{a}ngulo de d\'{e}ficit promedio efectivo por
unidad de volumen.

\backmatter
\appendix

\bigskip

\chapter{\bigskip Ap\'{e}ndice 1: Variaci\'{o}n de la Acci\'{o}n de Regge}

En este Ap\'{e}ndice mostraremos como para una variedad $n-$dimensional
$M_{n}$ , se cumple que para variaciones infinitesimales de $l$ :

\bigskip%

\begin{equation}
\underset{p}{\sum}L_{p}^{n}\delta\varepsilon_{p}=0
\end{equation}

\bigskip donde $L_{p}^{n}$ es la medida $(n-2)-$dimensional del $p-$esimo hueso.

\bigskip

Sea $T_{n}$ un simplex, y definamos en el un conjunto de coordenadas
cartesianas. $T_{n}$ posee $n+1$ simplex frontera denotados $T_{n-1}^{r}$
$(r=1,2,...,n+1)$. Dos de estos simplex $T_{n-1}^{r}$ y $T_{n-1}^{s}$
tendr\'{a}n una frontera com\'{u}n denotada por $T_{n-2}^{rs}$.

Sea $U_{\mu}^{r}$ el vector unitario normal al $T_{n-1}^{r}$
\ $(r=1,2,...,n+1)$, \ es decir $\left(  U^{r}\right)  _{\mu}\left(
U^{r}\right)  ^{\mu}=1$. Si $\theta_{rs}$ es el \'{a}ngulo entre $T_{n-1}^{r}%
$\ y $T_{n-1}^{s}$ entonces tenemos:

\bigskip%

\begin{equation}
\cos\theta_{rs}=\left(  U^{r}\right)  _{\mu}\left(  U^{s}\right)  ^{\mu}
\label{ap3}%
\end{equation}

\bigskip

Ahora consideremos el tensor Sesqui- Sim\'{e}trico $\left(  U^{rs}\right)
_{\mu v}=-\left(  U^{rs}\right)  _{v\mu}$ \ \ definido por:

\bigskip%

\begin{equation}
\sin\left(  \theta_{rs}\right)  \left(  U^{rs}\right)  _{v\mu}=\left(
U^{r}\right)  _{\mu}\left(  U^{s}\right)  _{v}-\left(  U^{r}\right)
_{v}\left(  U^{s}\right)  _{\mu} \label{ap2}%
\end{equation}

\bigskip

Claramente tenemos que $\left(  U^{rs}\right)  _{\mu v}\left(  U^{rs}\right)
^{\mu v}=2$ , por lo que este tensor es el mismo que se utiliz\'{o} en la
Secci\'{o}n \ref{variacion2} para calcular el Tensor de Riemann. Este vector
define la orientaci\'{o}n de $T_{n-2}^{rs}$.

\bigskip

Ahora bien, si $L_{rs}$ es la medida del simplex $T_{n-2}^{rs}$ tenemos la identidad:

\bigskip%

\begin{equation}
\underset{s}{\sum}L_{rs}U^{rs}=0 \label{ap1}%
\end{equation}

\bigskip

Para comprobar \ref{ap1} tomemos por ejemplo un sistema en el que $\left(
U^{r}\right)  _{\mu}=1$ y $\left(  U^{r}\right)  _{\mu\neq0}^{{}}=0$. Entonces
las \'{u}nicas componentes no nulas de $\left(  U^{rs}\right)  _{v\mu}$ son
las que tengan $v\neq0$, y por lo tanto llamamos $\left(  V^{s}\right)
_{v}^{{}}=\left(  U^{rs}\right)  _{v0}$. Tenemos entonces $\left(
V^{s}\right)  _{v}\left(  V^{s}\right)  ^{v}=1$. Ahora bien, en $T_{n-1}^{r}$,
la cantidad $\underset{s}{\sum}L_{rs}\left(  V^{s}\right)  _{v}A^{v}$ es
simplemente el flujo del vector constante $A^{v}$ a trav\'{e}s de la frontera
de $T_{n-1}^{r}$ y por lo tanto debe ser cero. As\'{i} llegamos a la
ecuaci\'{o}n \ref{ap1}.

\bigskip

Diferenciando la ecuaci\'{o}n \ref{ap3}\ tenemos:

\bigskip%

\begin{equation}
\delta\theta_{rs}=-\frac{1}{\sin\theta_{rs}}\left[  \left(  U^{r}\right)
_{\mu}\delta\left(  U^{s}\right)  ^{\mu}+\left(  U^{s}\right)  _{\mu}%
\delta\left(  U^{r}\right)  ^{\mu}\right]  \label{ap4}%
\end{equation}

\bigskip

Por otro lado, multiplicando \ref{ap2} por $\left(  U^{r}\right)  ^{v}$ obtenemos:

\bigskip%

\begin{align}
\left(  U^{rs}\right)  _{v\mu}\left(  U^{r}\right)  ^{v}  &  =\frac{1}%
{\sin\theta_{rs}}\left[  \left(  U^{r}\right)  _{\mu}\left(  U^{s}\right)
_{v}\left(  U^{r}\right)  ^{v}-\left(  U^{r}\right)  _{v}\left(  U^{r}\right)
^{v}\left(  U^{s}\right)  _{\mu}\right] \\
\left(  U^{rs}\right)  _{v\mu}\left(  U^{r}\right)  ^{v}  &  =\frac{1}%
{\sin\theta_{rs}}\left[  \cos\theta_{rs}\left(  U^{r}\right)  _{\mu}-\left(
U^{s}\right)  _{\mu}\right]
\end{align}

\bigskip

\bigskip Multiplicando esta \'{u}ltima expresi\'{o}n por $\delta\left(
U^{r}\right)  ^{\mu}$ y utilizando $\left(  U^{r}\right)  _{\mu}\delta\left(
U^{r}\right)  ^{\mu}=0$ tenemos:

\bigskip%
\begin{equation}
\left(  U^{rs}\right)  _{v\mu}\left(  U^{r}\right)  ^{v}\delta\left(
U^{r}\right)  ^{\mu}=-\frac{1}{\sin\theta_{rs}}\left(  U^{s}\right)  _{\mu
}\delta\left(  U^{r}\right)  ^{\mu}%
\end{equation}

\bigskip

Reemplazando esta ecuaci\'{o}n en \ref{ap4} tenemos:

\bigskip%

\begin{equation}
\delta\theta_{rs}=\left(  U^{rs}\right)  _{v\mu}\left[  \left(  U^{r}\right)
^{v}\delta\left(  U^{r}\right)  ^{\mu}+\left(  U^{s}\right)  ^{v}\delta\left(
U^{s}\right)  ^{\mu}\right]
\end{equation}

\bigskip

Multipilcando por $L_{rs}$ y sumando obtenemos:

\bigskip%

\begin{equation}
\underset{r,s}{\sum}L_{rs}\delta\theta_{rs}=\underset{r,s}{\sum}L_{rs}\left(
U^{rs}\right)  _{v\mu}\left[  \left(  U^{r}\right)  ^{v}\delta\left(
U^{r}\right)  ^{\mu}+\left(  U^{s}\right)  ^{v}\delta\left(  U^{s}\right)
^{\mu}\right]
\end{equation}

\bigskip

Utilizando \ref{ap1} se tiene:

\bigskip%

\begin{equation}
\underset{r,s}{\sum}L_{rs}\delta\theta_{rs}=0
\end{equation}

\bigskip

Si sumamos esta expresi\'{o}n sobre todos los simplex del esqueleto obtenemos finalmente:

\bigskip%

\begin{equation}
\underset{p}{\sum}L_{p}^{n}\delta\varepsilon_{p}=0
\end{equation}

Es decir que la variaci\'{o}n de la Acci\'{o}n de Regge sera simplemente:%

\begin{equation}
\delta I=\frac{1}{8\pi}\underset{k}{\delta\sum}\varepsilon_{k}L_{k}=\frac
{1}{8\pi}\underset{k}{\sum}\left[  L_{k}\delta\varepsilon_{k}+\varepsilon
_{k}\delta L_{k}\right]  =\frac{1}{8\pi}\underset{k}{\sum}\varepsilon
_{k}\delta L_{k}=0
\end{equation}

\bigskip

\chapter{Ap\'{e}ndice 2: Cambio en el \'{a}rea de un tri\'{a}ngulo
en\label{apendice1} funci\'{o}n de la variaci\'{o}n en la longitud de sus lados.}

\bigskip

\begin{figure}
[h]
\begin{center}
\includegraphics[
natheight=109.062500pt,
natwidth=189.562500pt,
height=95.5625pt,
width=164.75pt
]%
{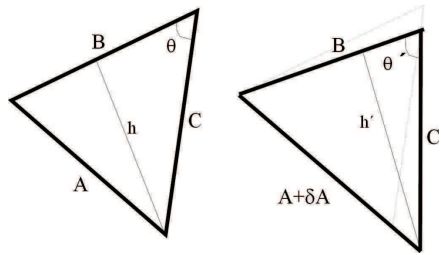}%
\caption{Cambio en el area de un triangulo}%
\label{areat}%
\end{center}
\end{figure}

Consideremos el triangulo $ABC$ de la Figura \ref{areat}. Si hacemos una
peque\~{n}a variaci\'{o}n $\delta A$ en uno de sus lados manteniendo
constantes los lados $B$ y $C$ obtendremos el triangulo $A%
\acute{}%
BC$ que se observa.

\bigskip%

El \'{a}rea del tri\'{a}ngulo $ABC$ ser\'{a} simplemente $\frac{1}{2}Bh$. Por
otro lado el \'{a}rea del tri\'{a}ngulo $A%
\acute{}%
BC$ sera $\frac{1}{2}Bh%
\acute{}%
$. De esta manera el cambio en el \'{a}rea es:

\bigskip%

\begin{equation}
\delta L=\frac{1}{2}B\left(  h%
\acute{}%
-h\right)
\end{equation}

\bigskip

De la Figura \ref{areat}\ \ observamos ademas que \ $h=C\sin\theta$ mientras
que $\ h%
\acute{}%
=C\sin\theta%
\acute{}%
$. Asi tenemos:

\bigskip%

\begin{equation}
\delta L=\frac{1}{2}BC\left(  \sin\theta%
\acute{}%
-\sin\theta\right)  \label{area1}%
\end{equation}

\bigskip

Utilizando el Teorema del Coseno tenemos $\ A^{2}=B^{2}+C^{2}-2BC\cos\theta$.
De donde, diferenciando, obtenemos:

\bigskip%
\begin{equation}
2A\delta A=2BC\sin(\theta)\delta\theta
\end{equation}%

\begin{equation}
\frac{A\delta A}{BC\sin\theta}=\delta\theta\label{area2}%
\end{equation}

\bigskip

Ahora bien, el \'{a}ngulo $\theta%
\acute{}%
$ ser\'{a} simplemente $\theta%
\acute{}%
=\theta+\delta\theta$. De esta manera tenemos que:

\bigskip%

\begin{equation}
\sin\theta%
\acute{}%
=\sin(\theta+\delta\theta)=\sin\theta\cos\delta\theta+\cos\theta\sin
\delta\theta
\end{equation}

\bigskip

Ya que estamos considerando $\delta A$ infinitesimal, tendremos tambien
$\delta\theta$ infinitesimal y por lo tanto podemos aproximar:

\bigskip%
\begin{equation}
\sin\theta%
\acute{}%
=\sin\theta\cos\delta\theta+\cos\theta\sin\delta\theta\approx\sin\theta
+\cos(\theta)\delta\theta
\end{equation}

\bigskip Reemplazando en la ecuaci\'{o}n \ref{area1} tenemos:

\bigskip%
\begin{equation}
\delta L=\frac{1}{2}BC\left(  \sin\theta+\cos(\theta)\delta\theta-\sin
\theta\right)  =\frac{1}{2}BC\cos(\theta)\delta\theta
\end{equation}

\bigskip Utilizando ahora la ecuaci\'{o}n \ref{area2} obtenemos:%

\begin{equation}
\delta L=\frac{1}{2}BC\cos\theta\left(  \frac{A\delta A}{BC\sin\theta}\right)
\end{equation}%

\begin{equation}
\delta L=\frac{1}{2}\cot(\theta)A\delta A
\end{equation}

\bigskip

Si ahora numeramos los lados del triangulo como $l_{1},l_{2},l_{3}$ y
realizamos variaciones a cada uno de ellos tendremos que la variaci\'{o}n en
el \'{a}rea del tri\'{a}ngulo ser\'{a}:

\bigskip%

\begin{equation}
\delta L=\frac{1}{2}\underset{p}{\sum}l_{p}\cot(\theta_{p})\delta l_{p}%
\end{equation}

\bigskip

donde $\theta_{p}$ es el \'{a}ngulo opuesto al lado $l_{p}$.

\bigskip

\

\end{document}